\documentstyle[12pt,epsf]{article}

\def\csumb{
$^{a}$ Institute of Theoretical Physics, Academia Sinica, Beijing 100080, China \\
$^{b}$ Ottawa-Carleton Institute for Physics,
Department of Physics, Carleton University, Ottawa, Canada K1S 5B6
}

\def\Title#1{\begin{center} {\Large\bf #1 } \end{center}}
\def\Author#1{\begin{center}{ \sc #1} \end{center}}
\def\Address#1{\begin{center}{ \it #1} \end{center}}

\def\vereq#1#2{\lower3pt\vbox{\baselineskip1.5pt \lineskip1.5pt
\ialign{$\m@th#1\hfill##\hfil$\crcr#2\crcr\sim\crcr}}}

\newcommand{\mw}{M_{\rm W}}
\def\as{\alpha_s}
\newcommand{\aem}{\alpha}

\newcommand{\bea}{\begin{eqnarray}}
\newcommand{\eea}{\end{eqnarray}}
\newcommand{\beq}{\begin{equation}}
\newcommand{\eeq}{\end{equation}}

\begin{document}

\renewcommand{\thefootnote}{\fnsymbol{footnote}}

\vskip .1in \Title{ CP violation in $B \to \Phi K_S$ in a model
III 2HDM. }

\vfill \Author{Chao-Shang Huang $^{a}$ \footnote{E-mail address:
csh@itp.ac.cn}
 and  Shou-hua Zhu $^{b}$ \footnote{E-mail address: huald@physics.carleton.ca}  }
\Address{\csumb}
\vfill
The mixing induced time dependent CP asymmetry, direct CP
asymmetry, and branching ratio in $B \to \Phi K_S$ in a model III
2HDM are calculated, in particular, neutral Higgs boson
contributions are included. It is shown that satisfying all the
relevant experimental constraints, for time dependent CP asymmetry
$S_{\phi K}$ the model III can agree with the present data,
$S_{\phi k}=-0.39\pm 0.41$, within the $1\sigma$ error, and the
direct CP asymmetry which is zero in SM can be about $ -8\%  \sim
-20\%$ in the reasonable regions of parameters.
\vfill


\eject \baselineskip=0.3in




\section{Introduction}
The recently reported measurements of time dependent CP
asymmetries in $B \to \Phi K_S$ decays \footnote{The 2003 new
results are: $S_{\phi k}= -0.96\pm 0.50^{+0.09}_{-0.11}$ by
Belle\cite{newbelle} and $+0.45\pm 0.43\pm 0.07$ by
BaBar\cite{newbabar}.}
 by BaBar \cite{babarphi}
\begin{eqnarray}
\label{eq:babar} \sin (2 \beta (\Phi K_S))_{BaBar}=-0.19
^{+0.52}_{-0.50} \pm 0.09
\end{eqnarray}
and Belle \cite{bellephi}
\begin{eqnarray}
\label{eq:belle} \sin (2 \beta (\Phi K_S))_{Belle}=-0.73 \pm 0.64
\pm 0.18
\end{eqnarray}
 result in the error weighted average
\begin{equation}
\label{eq:ave} \sin (2 \beta (\Phi K_S))_{ave}=-0.39 \pm 0.41 \;
\end{equation}
with errors added in quadrature. The value in (\ref{eq:belle})
corresponds to the coefficient of the sine term in the time
dependent CP asymmetry \cite{Anikeev:2001rk}, see Section IV.
Belle also quotes  a value for the direct CP asymmetry $A_{CP}= -
C_{\Phi K}$, i.e., ~the cosine term, $C_{\Phi K}=-0.19 \pm 0.30$
\cite{bellephi,mon}. Although there are at present large
theoretical uncertainties in calculating strong phases, we still
examine direct CP asymmetry in the paper in order to obtain
qualitatively feeling for effects of new physics on CP violation.

In the SM the above asymmetry is related to that in $B \to J/\Psi
K_S$ \cite{Grossman:1996ke}-\cite{Grossman:1997gr} by
\begin{equation}
\label{eq:diff} \sin (2 \beta (\Phi K))= \sin (2 \beta (J/\Psi
K)\!) \! + \! O(\lambda^2 \!)
\end{equation}
where $\lambda \simeq 0.2$ appears in Wolfenstein's
parameterization of the CKM matrix and $\sin (2 \beta (J/\Psi
K_{S,L}))_{world-ave}=+0.734 \pm 0.054$.  Therefore,
(\ref{eq:ave})  violates the SM at the 2.7 $\sigma$ deviation.
The impact of these experimental results on the validity of CKM
and SM is currently statistics limited. Future prospects at the
$B$-factories are that the statistical error $\sigma_{\Phi
K_S}(stat)$ can be expected to improve roughly by a factor of
three with an increase of integrated luminosity from $0.1ab^{-1}$
to $1ab^{-1}$ \cite{Eigen:2001mk} and it will take some time
before we know the deviation with sufficient precision to draw
final conclusions.

However, the possibility of a would-be measurement of $\sin (2
\beta(\Phi K_S))=-0.39$  or a similar value which departs
drastically from the SM expectation of (\ref{eq:diff}) has
attracted much interest to search for new physics, in particular,
supersymmetry, two Higgs doublet model (2HDM), and
model-independent way \cite{sphik,chw}. In the paper we consider
the decay $B \to \Phi K_S$ in a model III 2HDM. It is well-known
that in the model III 2HDM the couplings involving Higgs bosons
and fermions can have complex phases, which can induce CP
violation effects, even in the simplest case in which all
tree-level FCNC couplings are negligible. The effect of the color
dipole operator on the phase from the decay amplitudes, $\Delta
\Phi\equiv arg(\frac{\bar A}{A})$, in $b\rightarrow s\bar s s$ in
the model III 2HDM has been studied in the second paper of
ref.~\cite{sphik} by Hiller and the result is $\Delta \Phi\leq
0.2$ which is far from explaining the deviation. We would like to
see if it is possible to explain the deviation in the model III
2HDM under all known experimental constraints by extending to
include the neutral Higgs boson (NHB) contributions and calculate
hadronic matrix elements to the $\alpha_s$ order. Some relevant
Wilson coefficients at the leading order (LO) in the model III
2HDM have been given~\cite{chao}. Because the hadronic matrix
elements of relevant operators have been calculated to the
$\alpha_s$ order~\cite{bbns}, we can obtain the amplitude of the
process to the $\alpha_s$ order if we know the relevant Wilson
coefficients at the next to leading order (NLO). In the paper we
calculate them at NLO in the model III 2HDM. Furthermore, as
pointed in ref.~\cite{chw} the NHB penguin induced operators
contribute sizably to both the branching ratio (Br) and time
dependent CP asymmetry $S_{\phi k}$ in supersymmetrical models. In
the paper we calculate the Wilson coefficients of NHB penguin
induced operators in the model III 2HDM. Our results show that in
the model III 2HDM, the CP asymmetry $S_{\phi K}$ can agree with
the present data, $S_{\phi k}=-0.39\pm 0.41$, within the $1\sigma$
error. Even if the $S_{\phi k}$ is measured to a level of $-0.4\pm
0.1$ in the future, the model III can still agree with the data at
the $2\sigma$ level. And the direct CP asymmetry can reach about
$-20\%$.

The paper is organized as follows. In section II we describe the
model III 2HDM briefly. In section III we  give the effective
Hamiltonian responsible for $B\rightarrow \phi K_s$ in the model.
In particular, we give the Wilson coefficients at NLO for the
operators which exist in SM and at LO for the new operators which
are induced by NHB penguins respectively. We present the decay
amplitude and the CP asymmetry $S_{\phi K}$ in $B\rightarrow \phi
K_s$ in Section IV. The Section V is devoted to numerical results.
In Section VI we draw our conclusions and present some
discussions.


\section{Model III two-Higgs-doublet model (2HDM)}

In model III 2HDM, both the doublets can couple to the up-type and
down-type quarks, the details of the model can be found in Ref.
\cite{modelIII}. The Yukawa Lagrangian relevant to our discussion
in this paper is
\begin{eqnarray}
{\cal L}_Y &=&
   - \frac{g}{2M_W} (H^0\cos\alpha - h^0 \sin\alpha)
     \biggr(\overline{U} M_U U + \overline{D} M_D D \biggr ) \nonumber \\
&-&
\frac{H^0 \sin\alpha + h^0\cos\alpha }{\sqrt{2}} \Biggr[
  \overline{U} \biggr( {\hat\xi}^U \hbox{$1\over2$}(1+\gamma^5) + {\hat\xi}^{U\dagger}
 \hbox{$1\over2$}(1-\gamma^5)
         \biggr ) U  \nonumber \\
&& +\overline{D} \biggr( {\hat\xi}^D \hbox{$1\over2$}(1+\gamma^5) +
 {\hat\xi}^{D\dagger} \hbox{$1\over2$}(1-\gamma^5)
         \biggr ) D
  \Biggr ]  \nonumber \\
&+& \frac{i A^0}{\sqrt{2}} \Biggr [
    \overline{U} \biggr( {\hat\xi}^U
           \hbox{$1\over2$}(1+\gamma^5) -{\hat\xi}^{U\dagger}
           \hbox{$1\over2$}(1-\gamma^5)
        \biggr ) U
   -\overline{D} \biggr( {\hat\xi}^D
           \hbox{$1\over2$}(1+\gamma^5) - {\hat\xi}^{D\dagger}
           \hbox{$1\over2$}(1-\gamma^5)
        \biggr ) D
     \Biggr ] \nonumber \\
&& -  H^+ \overline{U} \biggr[ V_{\rm CKM} {\hat \xi}^D
             \hbox{$1\over2$}(1+\gamma^5) -
   {\hat\xi}^{U\dagger} V_{\rm CKM}
             \hbox{$1\over2$}(1-\gamma^5) \biggr] D
  \nonumber \\
&&-  H^- \overline{D} \biggr[ {\hat \xi}^{D\dagger} V^\dagger_{\rm CKM}
   \hbox{$1\over2$}(1-\gamma^5) -
      V^\dagger_{\rm CKM} {\hat\xi}^U \hbox{$1\over2$}(1+\gamma^5) \biggr] U
 \;\;,
\end{eqnarray}
where $U$ represents the mass eigenstates of $u,c,t$ quarks and
$D$ represents the mass eigenstates of $d,s,b$ quarks, $V_{\rm
CKM}$ is the Cabibbo-Kobayashi-Maskawa matrix and the FCNC
couplings are contained in the matrices $\hat\xi^{U,D}$. The
Cheng-Sher ansatz for $\hat\xi^{U,D}$ is \cite{modelIII}
\begin{equation}
\label{anat}
\hat\xi^{U,D}_{ij} = \lambda_{ij} \frac{g\sqrt{m_i m_j}}{\sqrt{2}M_W}
\end{equation}
by which the quark-mass hierarchy ensures that the FCNC within the
first two generations are naturally suppressed by the small quark
masses, while a larger freedom is allowed for the FCNC involving
the third generations\footnote{Model III 2HDM  has a remarkably
stable FCNC suppression when one evolves the FCNC Yukawa coupling
parameters by the RGE's to higher energies\cite{Cvetic:1998uw}.}.
In the ansatz the residual degree of arbitrariness of the FC
couplings is expressed through the $\lambda_{ij}$ parameters which
are of order one and need to be constrained by the available
experiments. In the paper we choose $\xi^{U,D}$ to be diagonal and
set the u and d quark masses to be zero for the sake of simplicity
so that besides Higgs boson masses only $\lambda_{ii}$, i=s, c, b,
t, are the new parameters and will enter into the Wilson
coefficients relevant to the process.

\section{Effective Hamiltonian}

The effective Hamiltonian for charmless B decays with $\Delta B =
1$ is given by \cite{chw,burasbeast}

\begin{eqnarray}\label{eff}
 {\cal H}_{\rm eff} &=& \frac{G_F}{\sqrt2} \sum_{p=u,c} \!
   V_{pb} V^*_{ps} \bigg( C_1\,Q_1^p + C_2\,Q_2^p
   + \!\sum_{i=3,\dots, 16}\![ C_i\,Q_i+ C_i^\prime\,Q_i^\prime]
   \nonumber \\&& + C_{7\gamma}\,Q_{7\gamma}
   + C_{8g}\,Q_{8g}
   + C_{7\gamma}^\prime\,Q_{7\gamma}^\prime
   + C_{8g}^\prime \,Q_{8g}^\prime \, \bigg) + \mbox{h.c.} \,
\end{eqnarray}
Here $Q_i$ are quark and gluon operators and are given by
\begin{eqnarray}
&&Q_1^p = (\bar s_\alpha p_\beta)_{V-A} (\bar p_\beta
b_\alpha)_{V-A},\;\;
Q_2^p = (\bar s_\alpha p_\alpha)_{V-A} (\bar p_\beta b_\beta)_{V-A},\nonumber\\
&&Q_{3(5)} = (\bar s_\alpha b_\alpha)_{V-A}\sum_{q} (\bar q_\beta
q_\beta)_{V-(+)A},\;\; Q_{4(6)} = (\bar s_\alpha
b_\beta)_{V-A}\sum_{q}
(\bar q_\beta q_\alpha)_{V-(+)A},\nonumber\\
&&Q_{7(9)} = {3\over 2}(\bar s_\alpha b_\alpha)_{V-A}\sum_{q}
e_{q}(\bar q_\beta q_\beta)_{V+(-)A},\;\; Q_{8(10)} ={3\over 2}
(\bar s_\alpha b_\beta)_{V-A}\sum_{q}
e_{q}(\bar q_\beta q_\alpha)_{V+(-)A},\nonumber\\
&&Q_{11(13)} = (\bar s\, b)_{S+P}
\sum_q\,{m_q\lambda_{qq}^{*}(\lambda_{qq})\over m_b} (\bar q\,
q)_{S-(+)P} \,, \nonumber\\
&&Q_{12(14)} = (\bar s_i \,b_j)_{S+P}
 \sum_q\,{m_q\lambda_{qq}^{*}(\lambda_{qq})\over m_b}(\bar q_j \,q_i)_{S-(+)P} \,, \nonumber\\
&&Q_{15} = \bar s \,\sigma^{\mu\nu}(1+\gamma_5) \,b
\sum_q\,{m_q\lambda_{qq}\over m_b}
    \bar q\, \sigma_{\mu\nu}(1+\gamma_5)\,q \,, \nonumber\\
&&Q_{16} = \bar s_i \,\sigma^{\mu\nu}(1+\gamma_5) \,b_j \sum_q\,
    {m_q\lambda_{qq}\over m_b} \bar q_j\, \sigma_{\mu\nu}(1+\gamma_5) \,q_i
    \, \nonumber \\
&&Q_{7\gamma} = {e\over 8\pi^2} m_b \bar s_\alpha \sigma^{\mu\nu}
F_{\mu\nu}
(1+\gamma_5)b_\beta, \nonumber\\
&&Q_{8g} = {g_s\over 8\pi^2} m_b \bar s_\alpha \sigma^{\mu\nu}
G_{\mu\nu}^a {\lambda_a^{\alpha \beta}\over 2}(1+\gamma_5)b_\beta,
\end{eqnarray}
where $(V\pm A)(V\pm A) =\gamma^\mu(1\pm\gamma_5)
 \gamma_\mu(1\pm \gamma_5)$, $(\bar q_1 q_2)_{S\pm P}=\bar
 q_1(1\pm\gamma_5)q_2$,
$q = u,d,s,c,b$, $e_{q}$ is the electric charge number of $q$
quark, $\lambda_a$ is the color SU(3) Gell-Mann matrix, $\alpha$
and $\beta$ are color indices, and $F_{\mu\nu}$ [$G_{\mu\nu}$] are
the photon [gluon] field strengths. The operators $Q_i^\prime$s
are obtained from the unprimed operators $Q_i$s by exchanging
$L\leftrightarrow R$. In eq.(\ref{eff}) operators $Q_i$, i11,...,16, are induced by neutral Higgs boson
mediations~\cite{chw}.

The Wilson Coefficients $C_i$, i=1,...,10, have been calculated at
LO~\cite{burasbeast,chao}. We calculate them at NLO in the NDR
scheme and results are as follows.
\begin{eqnarray}
C_1(\mw) &=&     \frac{11}{2} \; \frac{\as(\mw)}{4\pi} \, ,
\label{eq:CMw1} \\
C_2(\mw) &=& 1 - \frac{11}{6} \; \frac{\as(\mw)}{4\pi}
               - \frac{35}{18} \; \frac{\aem}{4\pi} \, ,
\label{eq:CMw2} \\
C_3(\mw) &=& -\frac{\as(\mw)}{24\pi} \left\{ \widetilde{E}_0(x_t)+ E^{III}_0(y)\right\}
             +\frac{\aem}{6\pi} \frac{1}{\sin^2\theta_W}
             \left[ 2 B_0(x_t) + C_0(x_t) \right] \, ,
\label{eq:CMw3} \\
C_4(\mw) &=& \frac{\as(\mw)}{8\pi}\left\{  \widetilde{E}_0(x_t)+E^{III}_0(y) \right\} \, ,
\label{eq:CMw4} \\
C_5(\mw) &=& -\frac{\as(\mw)}{24\pi}\left\{ \widetilde{E}_0(x_t)+E^{III}_0(y) \right\} \, ,
\label{eq:CMw5} \\
C_6(\mw) &=& \frac{\as(\mw)}{8\pi} \left\{ \widetilde{E}_0(x_t)+E^{III}_0(y) \right\} \, ,
\label{eq:CMw6} \\
C_7(\mw) &=& \frac{\aem}{6\pi} \left[ 4 C_0(x_t) + \widetilde{D}_0(x_t)
\right]\, ,
\label{eq:CMw7} \\
C_8(\mw) &=& 0 \, ,
\label{eq:CMw8} \\
C_9(\mw) &=& \frac{\aem}{6\pi} \left[ 4 C_0(x_t) + \widetilde{D}_0(x_t) +
             \frac{1}{\sin^2\theta_W} (10 B_0(x_t) - 4 C_0(x_t)) \right] \, ,
\label{eq:CMw9} \\
C_{10}(\mw) &=& 0 \, , \\
C_{7\gamma}(M_W) &=& - \frac{A(x_t)}{2} - \frac{A(y)}{6} |\lambda_{tt}|^2
                     +B(y) \lambda_{tt} \lambda_{bb} e^{i\theta}  \;,\label{c7mw} \\
C_{8G} (M_W) &=& -\frac{D(x_t)}{2} - \frac{D(y)}{6} |\lambda_{tt}|^2 +
                     E(y) \lambda_{tt}\lambda_{bb} e^{i\theta} \label{c8mw} \;\;,
\end{eqnarray}
where $x_t = m_t^2/M_W^2$, and $y = m_t^2 / M_{H^\pm}^2$. Here the
Wilson coefficients $C_{7\gamma}$ and $C_{8g}$ at LO which are
given in ref.~\cite{chao} have also been written. The Wilson
coefficients $C_{7\gamma}$ and $C_{8g}$ at NLO in SM have been
given but they at NLO in model III 2HDM have not been calculated
yet. Because we calculate the decay amplitude only to the
$\alpha_s$ order it is enough to know them at LO. Here
\begin{eqnarray}
A(x) &=& x \biggr [ \frac{8x^2 +5x -7}{12(x-1)^3} - \frac{(3x^2 -2x)\ln x}
                           {2(x-1)^4} \biggr ] \\
B(y)&=&y\biggr[\frac{5y-3}{12 (y-1)^2} - \frac{(3y-2) \ln y}{6(y-1)^3}\biggr ]
\\
D(x)&=& x\biggr[ \frac{x^2 - 5x -2}{4(x-1)^3} + \frac{3x\ln x}{2(x-1)^4}\biggr
] \\
E(y)&=& y\biggr[\frac{y-3}{4(y-1)^2} + \frac{\ln y}{2(y-1)^3} \biggr ] \;\;
\\
B_0(x_t) &=& \frac{1}{4} \left[ \frac{x_t}{1-x_t}
+ \frac{x_t \ln x_t}{(x_t-1)^2}
\right]\, , \label{eq:Bxt} \\
C_0(x_t) &=& \frac{x_t}{8} \left[\frac{x_t-6}{x_t-1}
+ \frac{3 x_t + 2}{(x_t-1)^2}
\ln x_t \right]\, ,
\label{eq:Cxt} \\
D_0(x_t) &=& -\frac{4}{9} \ln x_t +
\frac{-19 x_t^3 + 25 x_t^2}{36 (x_t-1)^3} +
\frac{x_t^2 (5 x_t^2 - 2 x_t - 6)}{18 (x_t-1)^4} \ln x_t \, ,
\label{eq:Dxt} \\
\widetilde{D}_0(x_t) &=& D_0(x_t) - \frac{4}{9} \, .
\end{eqnarray}
and
\begin{eqnarray}
E_0(x_t) &=&
-\frac{2}{3} \ln x_t + \frac{x_t (18 -11 x_t - x_t^2)}{12 (1-x_t)^3} +
          \frac{x_t^2 (15 - 16 x_t  + 4 x_t^2)}{6 (1-x_t)^4} \ln x_t \, ,
\label{eq:Ext1} \\
\widetilde{E}_0(x_t) &=& E_0(x_t) - \frac{2}{3}
\\
E^{III}_0(y) &=&
 |\lambda_{tt}|^2
\left\{ \frac{16y-29y^2+7y^3}{36(1-y)^3}+\frac{2y-3y^2}{6(1-y)^4} \ln y \right\}.
\label{eq:Exttilde1}
\end{eqnarray}

The Wilson coefficients $C_i$, i=11,...,16, at the leading order
can be obtained from $C_{Q1}$ and $C_{Q2}$ in Ref. \cite{dhll}.
The non-vanishing coefficients at $m_W$ are
\begin{eqnarray}
C_{11} (M_W) &=& \frac{\alpha}{4 \pi} \frac{m_b}{m_\tau
\lambda_{\tau\tau}^*}
\left( C_{Q1}-C_{Q2} \right) \nonumber \\
C_{13} (M_W) &=& \frac{\alpha}{4 \pi} \frac{m_b}{m_\tau
\lambda_{\tau\tau}}
\left( C_{Q1}+C_{Q2} \right).
\end{eqnarray}
We shall omitted the contributions of the primed operators in
numerical calculations for they are suppressed by $m_s\over m_b$
in model III 2HDM.

For the process we are interested in this paper, the Wilson
coefficients should run to the scale of $O(m_b)$. $C_1-C_{10}$ are
expanded to $O(\alpha_s)$ and NLO RGEs should be used. However for
the $C_{8g}$ and $C_{7\gamma}$, LO results should be sufficient.
The details of the running of these Wilson coefficients can be
found in Ref. \cite{burasbeast}. The one loop anomalous dimension
matrices of the NHB induced operators can be divided into two
distangled groups~\cite{adm}
\begin{eqnarray}
\gamma^{(RL)}=\begin{tabular}{c|cccc} &$O_{11}$&$O_{12}$\\\hline
$O_{11}$&$-16$&0\\
$O_{12}$&-6&$2$
\end{tabular}
\end{eqnarray}
and
\begin{eqnarray}
\gamma^{(RR)}=\begin{tabular}{c|cccc}
&$O_{13}$&$O_{14}$&$O_{15}$&$O_{16}$\\\hline
$O_{13}$&$-16$&0&1/3&-1\\
$O_{14}$&-6&$2$&-1/2&-7/6\\
$O_{15}$&16&-48&$16/3$&0\\
$O_{16}$&-24&-56&6&$-38/3$
\end{tabular}
\end{eqnarray}
For $Q_i^\prime$ operators we have
\begin{eqnarray}
\gamma^{(LR)}=\gamma^{(RL)}~~~~~and~~~~~
\gamma^{(LL)}=\gamma^{(RR)}\,.
\end{eqnarray}
Because at present no NLO Wilson coefficients $C_i^{(\prime)}$,
i=11,...,16, are available we use the LO running of them in the
paper.

\section{The decay amplitude and CP asymmetry in
$B_{d}^{0} \rightarrow \phi K_S$} We use the BBNS
approach~\cite{bbns1} to calculate the hadronic matrix elements of
operators. In the approach the hadronic matrix element of a
operator in the heavy quark limit can be written as
\begin{eqnarray}
<\phi K|Q|B> = <\phi K|Q|B>_{f} [1+ \sum r_n \alpha_s^n ],
\end{eqnarray}
where $<\phi K|Q|B>_{f}$ indicates the naive factorization result.
The second term in the square bracket indicates higher order
$\alpha_s$ corrections to the matrix elements~\cite{bbns1}. We
calculate hadronic matrix elements to the $\alpha_s$ order in the
paper. In order to see explicitly the effects of new operators in
the model III 2HDM we divide the decay amplitude into two parts.
One has the same form as that in SM, the other is new. That is, we
can write the decay amplitude, to the $\alpha_s$ order, for $B\to
\phi K$ in the heavy quark limit as \cite{bbns,He:2000rf,chw}
\begin{eqnarray}
&&A(B\to \phi K)= {G_F\over \sqrt{2}} A
<\phi|\bar s\gamma_\mu s |0> <K|\bar s \gamma^\mu b|B>,\nonumber\\
&& A= A^o+A^n, \\&&A^o= V_{ub}V^*_{us}[a_3^u +a_4^u+a_5^u -
{1\over 2}(a_7^u + a_9^u + a_{10}^u
)+ a_{10a}^u] \nonumber\\
&&\;\;\;\;\;\; \; +V_{cb}V^*_{cs}[a_3^c +a_4^c+a_5^c - {1\over
2}(a_7^c + a_9^c + a_{10}^c )+ a_{10a}^c)], \label{ao}\\
&&A^n=-V_{tb}V^*_{ts}\bigg( a_4^{neu}+ {m_s \over m_b}[ -{1\over
2}\lambda_{ss}^{*}(a_{12}+a_{12}^\prime) + \lambda_{ss}\,{4
m_s\over m_b}\,(a_{16}+a_{16}^\prime) ]\bigg)\,.\label{an}
\end{eqnarray}
For the hadronic matrix elements of the vector current, we can
write $<\phi| \bar s \gamma_\mu b | 0 > = m_\phi f_\phi
\epsilon^\phi_\mu$ and $<K|\bar s \gamma^\mu b|B> = F_1^{B\to
K}(q^2) (p_B^\mu + p_K^\mu) +(F^{B\to K}_0(q^2)-F^{B\to K}_1(q^2))
(m_B^2-m_K^2)q^\mu/q^2$. Here, the coefficients $a_i^{u,c}$ in
eq.(\ref{ao}) are given by\footnote{The explicit expressions of
the coefficients $a_i^{u,c}$ have been given in
ref.~\cite{He:2000rf}. Because there are minor errors in the
expressions in the paper and in order to make this paper
self-contained we reproduce them here, correcting the minor
errors.}
\begin{eqnarray}
&&a_3^u=a^c_3 = c_3 + {c_4\over N} + {\alpha_s\over 4 \pi}
{C_F\over N} c_4 F_\phi,\nonumber\\
&&a_4^p = c_4 + {c_3\over N} +{\alpha_s\over 4\pi} {C_F\over N}
\left [ c_3(F_\phi + \hat{G}_\phi(s_s) + \hat{G}_\phi(s_b))
+ c_2 \hat{G}_\phi(s_p)\right. \nonumber\\
 &&\;\;\;\;+ \left .
(c_4+c_6) \sum_{f=u}^b \tilde G_\phi(s_f) + c_{8g}G_{\phi,8g}\right ],
\nonumber\\
&&a_5^u=a_5^c = c_5 +{c_6\over N} + {\alpha_s \over 4 \pi}
{C_F\over N} c_6(-F_\phi - 12),\nonumber\\
&&a_7^u = a^c_7 = c_7 + {c_8\over N} + {\alpha_s \over 4\pi}
{C_F\over N} c_8(-F_\phi -12),\nonumber\\
&&a_9^u = a_9^c = c_9 +{c_{10}\over N} + {\alpha_s\over 4\pi}
{C_F\over N} c_{10}F_\phi,\nonumber\\
&&a_{10}^u = a_{10}^c = c_{10} +{c_{9}\over N} + {\alpha_s\over 4\pi}
{C_F\over N} c_{9}F_\phi,\nonumber\\
&&a_{10a}^p= {\alpha_s\over 4\pi} {C_F\over N} \left [
(c_8+c_{10}) {3\over 2} \sum _{f=u}^be_f \hat{G}_\phi(s_f) +c_9
{3\over 2} (e_s \hat{G}_\phi(s_s) + e_b \hat{G}_\phi(s_b))\right
],
\end{eqnarray}
where $p$ takes the values $u$ and $c$, $N=3$,
$C_F = (N^2-1)/2N$, and $s_f = m^2_f/m_b^2$,
\begin{eqnarray}
&&\hat{G}_\phi(s) = {2\over 3} + {4\over 3} \ln{m_b\over \mu}
-G_{\phi}(s),\nonumber\\&& G_{\phi}(s) - 4\int^1_0 dx \Phi_\phi(x) \int^1_0 du u(1-u) \ln[s-u(1-u)(1-x)],\nonumber\\
&&\tilde G_\phi(s) = \hat{G}_\phi(s) -{2\over 3}, \nonumber\\
&&G_{\phi,8g} = -2 G^0_\phi,~~~~G^0_\phi = \int_0^1 \frac{dx}{\bar
x}\, \Phi_{\phi}(x)\,\nonumber\\
&&F_\phi = -12 \ln{\mu \over m_b} - 18 + f^I_\phi + f^{II}_\phi,
\nonumber\\
&&f^I_\phi = \int^1_0 dx g(x) \phi_\phi(x),\;\;
g(x) =  3{1-2x\over 1-x} \ln x -3i\pi,\nonumber\\
&&f^{II}_\phi = {4\pi^2\over N}
{f_K f_B\over F_1^{B\to K}(0) m_B^2}
\int^1_0 dz {\phi_B(z)\over z} \int^1_0 dx {\phi_K(x)\over x}
\int^1_0dy {\phi_\phi(y)\over y},
\end{eqnarray}
where $\phi_i(x)$ are meson wave functions,
\begin{eqnarray}\label{phikp}
&&\phi_B(x) = N_B x^2(1-x)^2 Exp[-{m_B^2 x^2\over 2 \omega_B^2}],
\nonumber\\
&&\phi_{K,\phi}(x) = 6x(1-x),
\end{eqnarray}
with normalization factor $N_B$ satisfying $\int^1_0 dx \phi_B(x)
= 1$. Fitting various B decay data, $\omega_B$ is determined to be
0.4 GeV \cite{Keum:2000ph}. In eq.(\ref{phikp}) the asymptotic
limit of the leading-twist distribution amplitudes for $\phi$ and
K has been assumed.

The coefficients $a_i$ in eq. (\ref{an}) are
\begin{eqnarray}\label{ai}
\vspace{0.1cm}
   a_4^{neu} &=& {C_F\alpha_s\over 4\pi}\,{P_{\phi,2}^{neu}\over N_c}
\,,\nonumber\\
   a_{12} &=& C_{12} + {C_{11}\over N_c} \left[ 1
    + {C_F \alpha _s \over 4 \pi  } \left( -V^\prime
     - f^{II}_\phi \right) \right]\,,\nonumber\\
   a_{16} &=&  C_{16} + \displaystyle{\frac{C_{15}}{N_c} }\,,
\end{eqnarray}
where \begin{eqnarray} V^\prime &=& -12 \ln{\mu \over m_b} - 6 +
\int^1_0 dx g(\bar x) \phi_\phi(x),\;\;
\end{eqnarray}
and
\begin{eqnarray}\label{P2phi}
   P_{\phi,2}^{neu} &=&  -{1\over 2}
 ( C_{11} + C_{11}^\prime ) \nonumber\\
     &&\ \ \times  \left[ {m_s\lambda_{ss}^{*} \over m_b}  \left( {4\over 3} \ln {
m_b \over \mu } - G_\phi (0)\right)  +\lambda_{bb}^{*}\left( {
4\over 3} \ln { m_b
\over \mu } - G_ \phi (1) \right) \right] \nonumber \\
  && + ( C_{13} + C_{13}^\prime )\lambda_{bb} \left[ -2 \ln { m_b \over \mu }
       \,G^0_\phi - GF_ \phi (1) \right] \nonumber\\
  &&- 4 (C_{ 15} +C_{15}^\prime)\lambda_{bb} \left[ \left( -{1\over 2}
-2 \ln { m_b \over \mu }\right) G^0_\phi - GF_\phi (1) \right] \nonumber\\
&& - 8  (C_{16}+C_{16}^\prime)\bigg[\lambda_{bb} \bigg(
  -2 \ln { m_b\over \mu }\,G^0_\phi
 - GF_\phi(1)
 \bigg) \nonumber\\
 && + \lambda_{cc}
 \, \left( {m_c \over m_b } \right)  ^2 \,
  \left( - 2 \ln { m_b \over \mu }\,G^0_\phi - GF_ \phi(s_c )
  \right)\bigg]
\end{eqnarray}
In eq. (\ref{P2phi})
\begin{eqnarray}
    GF_\phi (s)&=& \int^1_0 dx { \Phi _\phi (x)\over \bar x}
\, GF(s-i\, \epsilon , \bar x)\,, \nonumber\\
GF(s,x) &=& \int^1_0 dt \ln \big[ s-x \,t {\bar t} \big] \,,
\end{eqnarray}
with $\bar x =1-x$. In calculations we have set $m_{u,d}=0$ and
neglected the terms which are proportional to $m_s^2/m_b^2$ in eq.
(\ref{P2phi}). We have included only the leading twist
contributions in eq. (\ref{ai}). In eq.(\ref{an}) $a_i^\prime$ is
obtained from $a_i$ by substituting the Wilson coefficients
$C_j^\prime$s for $C_j$s. In numerical calculations $a_i^\prime$
is set to be zero because we have neglected $C_j^\prime$s. We see
from eq. (\ref{P2phi}) that the new contributions to the decay
amplitude can be large if the coupling $\lambda_{bb}$ is
large due to the large contributions to the hadronic elements of
the NHB induced operators at the $\alpha_s$ order arising from
penguin contractions with b quark in the loop.

The decay rate can be obtained \cite{He:2000rf}
\begin{eqnarray}
\Gamma(B\to \phi K) = {G_F^2\over 32\pi} |A|^2 f_\phi^2 F_1^{B\to
K}(m^2_\phi)^2 m^3_B P^{3/2}_{K\phi},
\end{eqnarray}
where $P_{ij} = (1-m^2_i/m_B^2-m^2_j/m^2_B)^2 - 4 m^2_i
m_j^2/m_B^4$.


The time-dependent $CP$-asymmetry $S_{\phi K}$ is given by
\begin{equation}
a_{\phi K}(t) = -C_{\phi K} \cos(\Delta M_{B_d^{0}} t) + S_{\phi
K} \sin(\Delta M_{B_d^{0}}t),
\end{equation}
where
\begin{equation}
C_{\phi K} = \frac{1-|\lambda_{\phi k}|^2}{1+|\lambda_{\phi k}|^2}
\,, \ \ \ \ \ \ \ \ \ \ S_{\phi K} = \frac{2\,\mathrm{Im}
\lambda_{\phi k}}{1+|\lambda_{\phi k}|^2} \; .\label{defi}
\end{equation}
Here $\lambda_{\phi k}$ is defined as
\begin{eqnarray}
\lambda_{\phi k} &=&
\left(\frac{q}{p}\right)_B
\frac{\mathcal{A}(\overline{B} \rightarrow \phi K_S)}{\mathcal{A}(B \rightarrow \phi K_S)}.
\end{eqnarray}
The ratio $(q/p)_B$
 is nearly a pure phase. In SM $\lambda_{\phi k}= e^{i2\beta}+ \! O(\lambda^2
 \!)$. As pointed out in Introduction, the model III can give a
 phase to the decay which we call
$\phi^{\mathrm{III}}$. Then we have
\begin{equation}
\lambda = e^{i(2\beta+\phi^{\mathrm{III}})}
\frac{|\bar{\mathcal{A}}|}{|\mathcal{A}|} \ \ \ \Rightarrow
S_{\phi K} = \sin (2\beta+\phi^{\mathrm{III}})
\end{equation}
if the ratio $\frac{|\bar{\mathcal{A}}|}{|\mathcal{A}|}=1$. In
general the ratio in the model III is not equal to one and
consequently it has an effect on the value of $S_{\phi K}$, as can
be seen from eq. (\ref{defi}). Thus the presence of the phases in
the Yukawa couplings of the charged and neutral Higgs bosons can
alter the value of $S_{\phi K}$ from the standard model prediction
of $S_{\phi K}=\sin 2\beta_{J/\psi K} \sim 0.7$.

\section{Numerical analysis}

\subsection{Parameters input}

In our numerical calculations we will use the following values for the
relevant parameters:
$m_b = 4.8$ GeV, $m_c = 1.5$ GeV,
$m_t=175$ GeV,
$\Lambda^{(5)}=225$ MeV, $2 \times 10^{-4} < Br(B\rightarrow X_s\gamma)
<4.5 \times 10^{-4}$, $d_n<10^{-25}\hbox{e$\cdot$cm}$,
$f_\phi = 0.233$ GeV,
$f_K = 0.158$ GeV,
$f_B = 0.18 $ GeV, $F^{B\rightarrow K}_1(m_\phi)=0.3$.
The parameters for CKM are $s_{12}=0.2229$, $s_{13}=0.0036$,
$s_{23}=0.0412$ and $\delta_{13}=1.02$.

\subsection{Constraints from $B \rightarrow X_s\gamma$ and neutron
electric dipole moment (NEDM)}

It is shown in Ref. \cite{chao} that the most strictly constraints
come from $B \rightarrow X_s\gamma$ and neutron electric dipole
moment (NEDM). For completeness, we write the formulas as
following \cite{Kagan:1998bh}:
\begin{equation}
\frac{Br(B\to X_s\gamma)}{Br(B\to X_c e \bar \nu_e)} \frac{|V^*_{ts} V_{tb}|^2}{|V_{cb}|^2} \, \frac{6\alpha_{\rm em}}
 {\pi f(m_c/m_b)} |C_{7\gamma}(m_b)|^2 \;,
\end{equation}
where $f(z) = 1-8z^2 +8z^6 - z^8 -24z^4 \ln z$ and
$Br(b\to c e^- \bar \nu)=10.45 \%$.

The NEDM can be expressed as
\begin{equation}
d^g_n=10^{-25}\hbox{e$\cdot$cm}
\  \hbox{Im}(\lambda_{tt}  \lambda_{bb})
\left({\alpha(m_n)\over\alpha(\mu)}\right)^{1\over2}
\left({\xi_g\over 0.1}\right)
H\left({m_t^2\over M^2_{H^\pm}}\right) \ ,
\end{equation}
with
\begin{equation}
H(y)={3\over2}{y\over (1-y)^2} \left(y-3-{2\log y\over 1-y}\right) \ .
\end{equation}

\subsection{ Numerical results for $B\rightarrow K_s \phi$}

We have scanned the parameter space in model III, in the following
we will show the results for several specific parameters.

\begin{figure}
\epsfxsize=12 cm \centerline{\epsffile{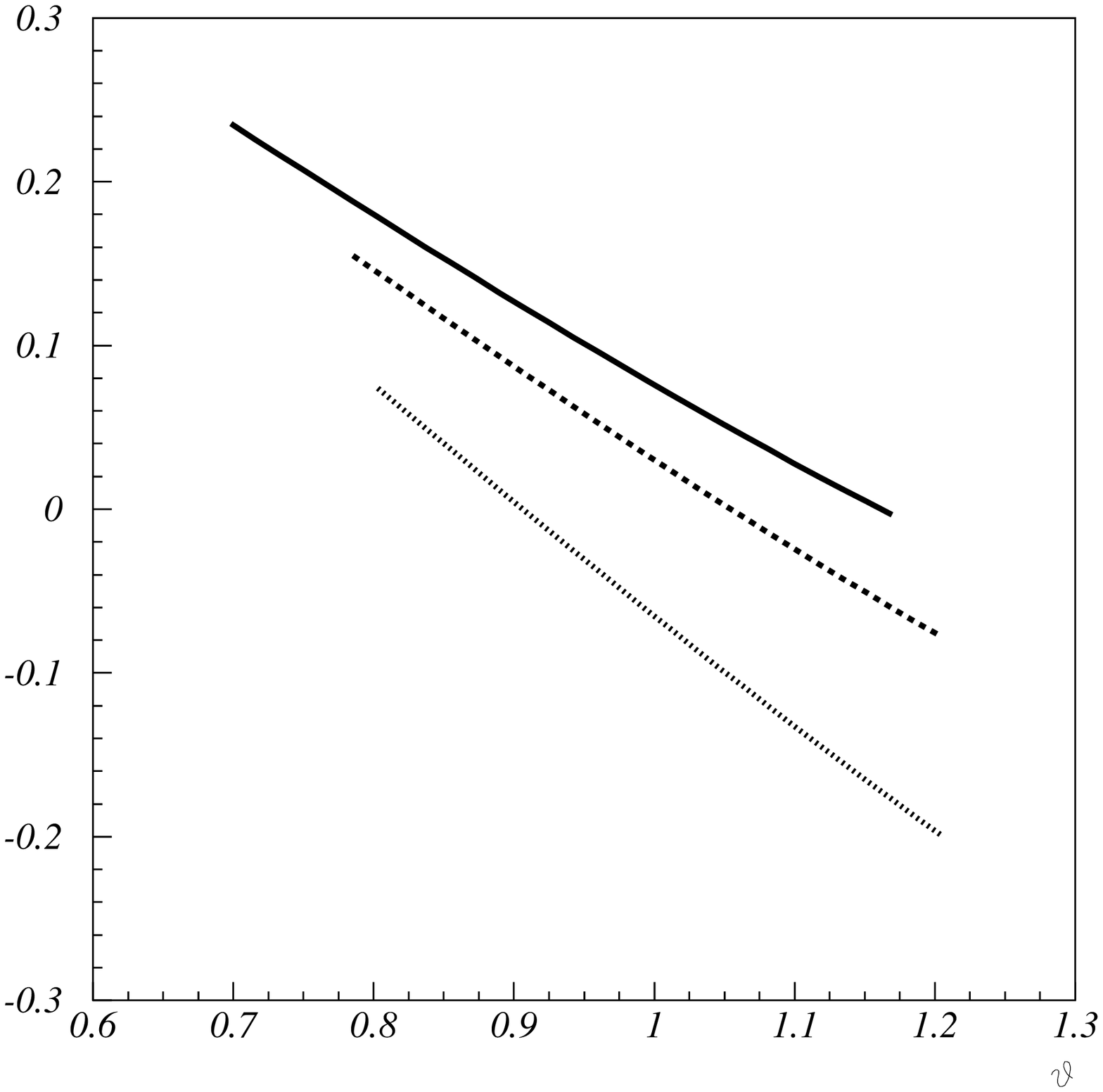}} \caption{
$S_{\phi K}$ as a function of $\theta \equiv \theta_{bb}+
\theta_{tt}$ with $\mu=2 m_b$ (solid), $m_b$ (dashed) and $m_b/2$
(dotted), where $m_{H^{\pm}}=$ 200 GeV, $|\lambda_{tt}|=0.03$,
$|\lambda_{bb}|=100$, $\lambda_{ss}=\lambda_{cc}=100 e^{-i
\pi/2}$. The parameter $\xi_g$ in neutron EDM expression is 0.03
\cite{chao,bu}. } \label{fig1}
\end{figure}

\begin{figure}
\epsfxsize=12 cm
\centerline{\epsffile{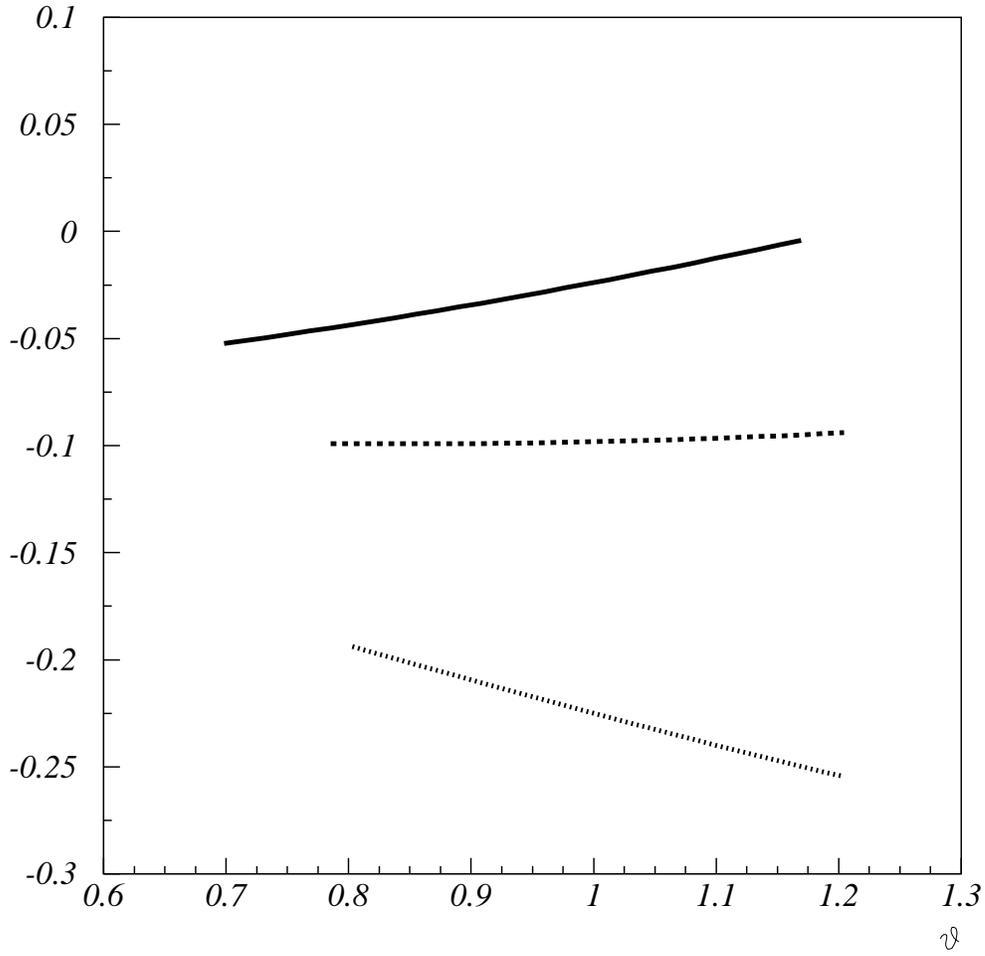}}
\caption{
$\Delta S$ [defined as difference between $S_{\phi K}$ with and
without NHB contributions]
as a function of $\theta$ with
$\mu=2 m_b$ (solid), $m_b$ (dashed) and $m_b/2$ (dotted).
 Other parameters are the same with Fig. \ref{fig1}.
}
\label{fig2}
\end{figure}

\begin{figure}
\epsfxsize=12 cm
\centerline{\epsffile{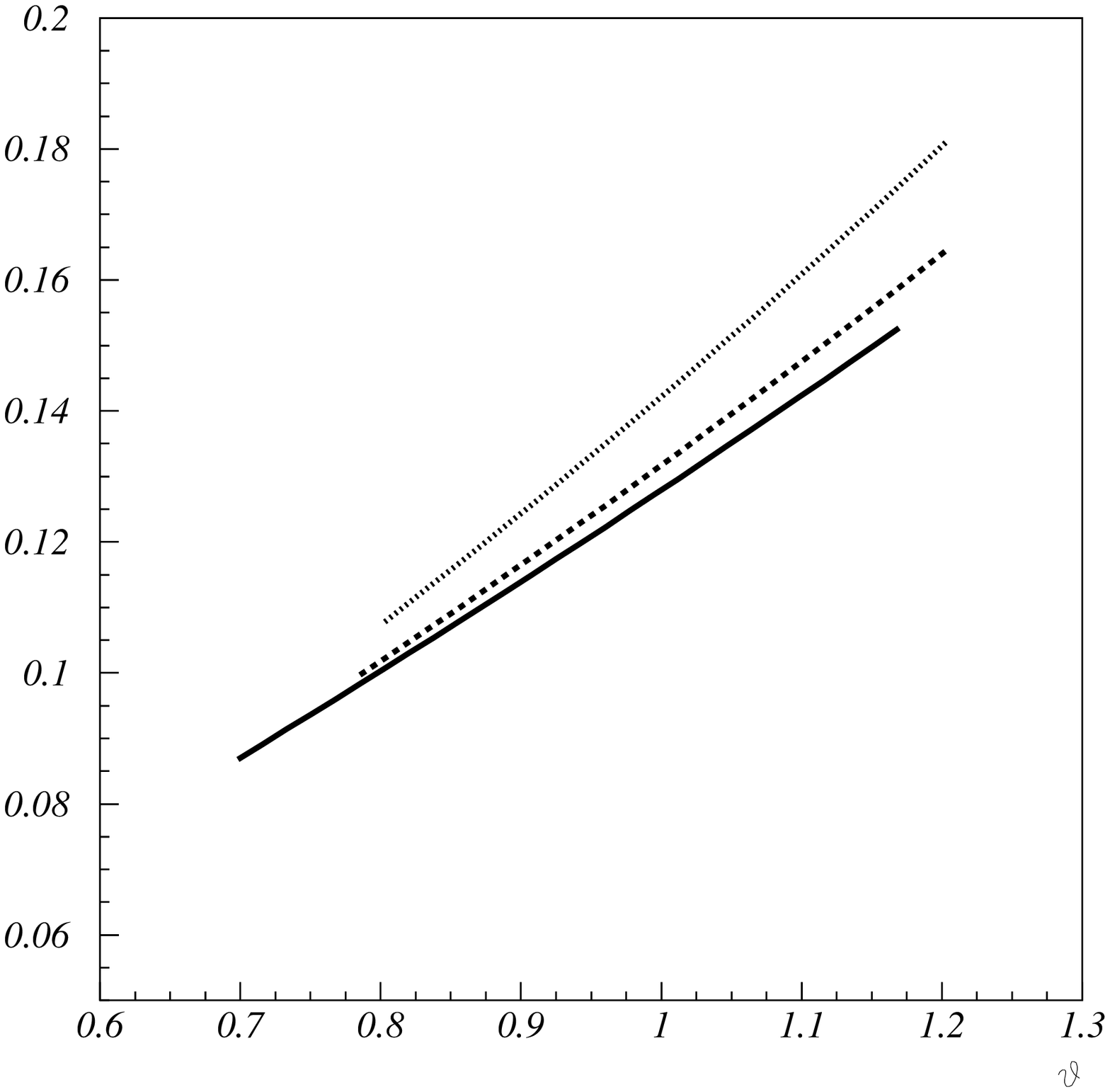}}
\caption{
$C_{\phi K}$ as a function of $\theta$. Other parameters and
conventions are the same with Fig. \ref{fig1}.}
\label{fig3}
\end{figure}

\begin{figure}
\epsfxsize=12 cm
\centerline{\epsffile{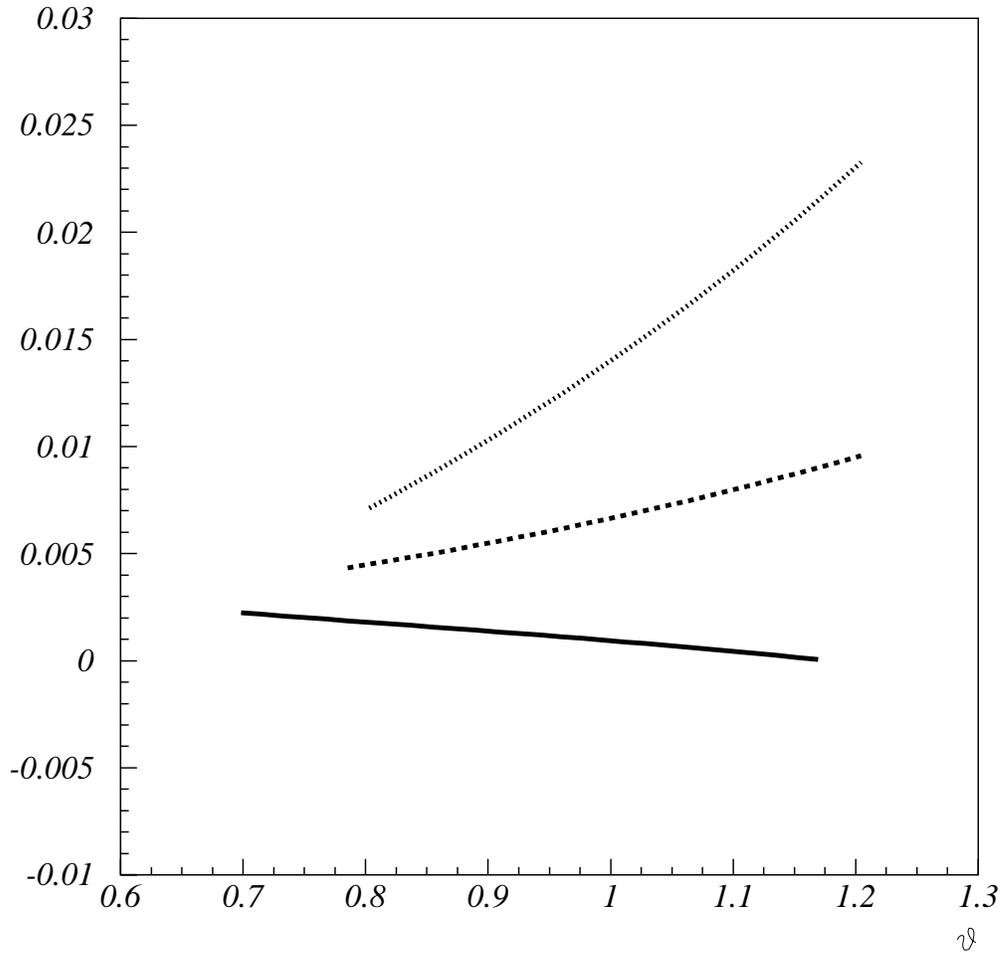}}
\caption{
$\Delta C$ [defined as difference between $C_{\phi K}$ with and
without NHB contributions]
as a function of $\theta$. Other parameters and
conventions are the same with Fig. \ref{fig3}.
}
\label{fig4}
\end{figure}

Figs. 1-4 are devoted to the case in which neutral Higgs boson
masses are set to be $m_{h^0}=115$ GeV, $m_{A^0}=120$ GeV,
$m_{H^0}=160$ GeV, which are the same with Ref. \cite{dhll}, and
consequently $C_{11}(m_W)\gg C_{13}(m_W)$. Figures \ref{fig1} and
\ref{fig2} show the $S_{\phi K}$ and $\Delta S$, defined as the
$S_{\phi K}$ difference with and without NHB contributions, as a
function of $\theta \equiv \theta_{bb}+\theta_{tt}$ with
$m_{H^{\pm}}=$ 200 GeV. Note that there is another allowed region
of $\theta$, about $-1.2 \sim -0.7$, in which $S_{\phi K}$ is
about 1. Therefore, we don't present the results in the figures.
From the figures we can see that in model III, the charged and
neutral Higgs boson contributions can decrease the value of
$S_{\phi K}$ down to -0.2, in the allowed parameter space. It
should be emphasized that the NHB contributions are sizable.
 In fig. \ref{fig3} and \ref{fig4}, we show the
direct CP violation variable $C_{\phi K}$
and  $\Delta C_{\phi
K}$, defined as $C_{\phi K}$ difference with and without NHB
contributions, as a function of $\theta$. It is obvious that
$C_{\phi K}$ can be 8-20 \%, i.e., it can be in agreement with the
data within $1\sigma$ deviation, while it is zero in the SM. At
the same time, the NHB contributions can only change $C_{\phi K}$
by less than 3\%.

Figs. 5-8 [and also in figs 9-11] are plotted for the case in
which the masses of NHBs have large splitting, $m_{A^0}=m_{H^0}=
1$ TeV $\gg m_{h^0}=115$ GeV, and consequently $C_{11}(m_W)$ is
the same order of magnitude, compared to $C_{13}(m_W)$. Now the
NHB contributions are as important as those of the charged Higgs
boson and $S_{\phi K}$ can reach about $-0.6$, as expected.

In order to demonstrate the NHB contributions, in figures
\ref{fig9}-\ref{fig11}, we show $S_{\phi K}$ as functions of the
phases of $\lambda_{bb}$ and $\lambda_{ss}$, $\theta_{bb}$ and
$\theta_{ss}$, and the correlation between $S_{\phi K}$ and
$C_{\phi K}$, respectively. It is clear that $S_{\phi K}$ is
sensitive to the phases.  At the same time, in the range [$-\pi,
\pi$] of $\theta_{bb}$ and $\theta_{ss}$ $C_{\phi K}$ changes only
several percents. There is a strong correlation between $S_{\phi
K}$ and $C_{\phi K}$ and $C_{\phi K}$ is always positive
regardless of the sign of $S_{\phi K}$, which is opposite to that
of the central value of measurements. Therefore, if the minus
$C_{\phi K}$ is confirmed in coming experiments the model III 2HDM
could be excluded.

\section{Conclusions and Discussions}

In summary we have calculated the Wilson coefficients at NLO for
the operators in the SM [except for $Q_{7\gamma}$ and $Q_{8g}$],
and at LO for the new operators which are induced by NHB penguins
in the model III 2HDM. Using the Wilson coefficients obtained, we
have calculated the mixing induced time-dependent CP asymmetry
$S_{\phi k}$, branching ratio and direct CP asymmetry $C_{\phi K}$
for the decay $B\rightarrow \phi K_s$. It is shown that in the
reasonable region of parameters where the constraints from
$B-\bar{B}$ mixing , $\Gamma(b \to s \gamma)$, $\Gamma(b \to c
\tau \bar{\nu}_\tau)$, $\rho_0$, $R_b$, $B\to \mu^+\mu^-$, and
electric dipole moments (EDMs) of the electron and neutron are
satisfied, the branching ratio of the decay can reach $10 \times
10^{-6}$, $C_{\phi K} $ can reach $18\%$ and $S_{\phi k}$ can be
negative in quite a large region of parameters and as low as -0.6
in some regions of parameters.

Let us separately discuss the two cases: 1) only the charged Higgs
contributions and 2) only the NHB contributions, in addition to
the SM ones. Without NHB contributions, i.e., in the first case,
the charged Higgs contributions can only decrease $S_{\phi k}$ to
around 0. That is, the model III can agree with the present data,
$S_{\phi k}=-0.39\pm 0.41$, within $1\sigma$ error.

For the second case, our results show that the effects of NHB
induced operators can be sizable even significant, depending on
the characteristic scale $\mu$ of the process. Due to the large
contributions to the hadronic elements of the operators at the
$\alpha_s$ order arising from penguin contractions with b quark in
the loop, both the Br and $S_{\phi K}$ are sizable or
significantly different from those in SM.

Putting all the contributions together, we conclude that the model
III can agree with the present data, $S_{\phi k}=-0.39\pm 0.41$,
within the $1\sigma$ error. Even if the $S_{\phi k}$ is measured
to a level of $-0.4\pm 0.1$ in the future, the model III can still
agree with the data at the $2\sigma$ level in quite a large
regions of parameters and at the $1\sigma$ level in some regions
of parameters. As for $C_{\phi K}$, our result is that it is
positive, which is opposite to that of the measured central value.
Considering the large uncertainties both theoretically and
experimentally at present, we should not take it seriously.

Our results show that both the Br and $S_{\phi k}$ (as well as
$C_{\phi K}$) of $B\to \phi K_S$ are sensitive to the
characteristic scale $\mu$ of the process, as can be seen from eq.
(\ref{P2phi}) and the SM amplitude. The significant scale
dependence comes mainly from the $O(\alpha_s)$ corrections of
hadronic matrix elements of the operators $Q_i$, i=11,...,16 and
also from leading order Wilson coefficients $C_i$, i=8g,11,...,16.
However, despite there is the scale dependence, the conclusion
that the model III can agree with the present data, $S_{\phi
k}=-0.39\pm 0.41$, within the $1\sigma$ error can still be drawn
definitely.
\section*{Note Added}We noticed the reference~\cite{gm} while
completing this work. In the ref.~\cite{gm} the mixing induced CP
asymmetry $S_{\phi K}$ in the model III 2HDM is investigated.
Comparing with the paper, our results on the Wilson coefficients
of the operators which exist in SM at NLO are in agreement. We
differ significantly from the paper in the neutral Higgs boson
contributions included. Furthermore, we calculate hadronic matrix
elements of operators to the $\alpha_s$ order by BBNS's approach
while the paper uses the naive factorization, i.e., at the tree
level. Therefore, our numerical results and consequently
conclusions are significantly different from those in the paper.
Even without including the NHB contributions our results are also
different from theirs due to the different precisions of
calculating hadronic matrix elements, to which $S_{\phi k}$ is
sensitive.

When the publication processing the paper, we were aware of Ref.
\cite{27} in which the LO anomalous dimensions for the mixing of
$Q_{11,12}$ onto $Q_{3,...,6}$ and $Q_9$ are given and those for
the mixing of $Q_{13,...,16}$ onto $Q_{7\gamma,8g}$ given in Ref.
\cite{28} are confirmed. In this paper these mixings are not taken
into account. If including them the numerical results would change
but the qualitative features of the results would be the same. We
shall include them in a forthcoming paper on``CP asymmetries in
$B\to \eta^\prime K_S$ and $\phi K_S$ in a model III 2HDM".

\section*{Acknowledgements}
This work was supported in part by the National Nature Science
Foundation of China, the Nature Sciences and Engineering Research
Council of Canada.



\begin{figure}
\epsfxsize=12 cm
\centerline{\epsffile{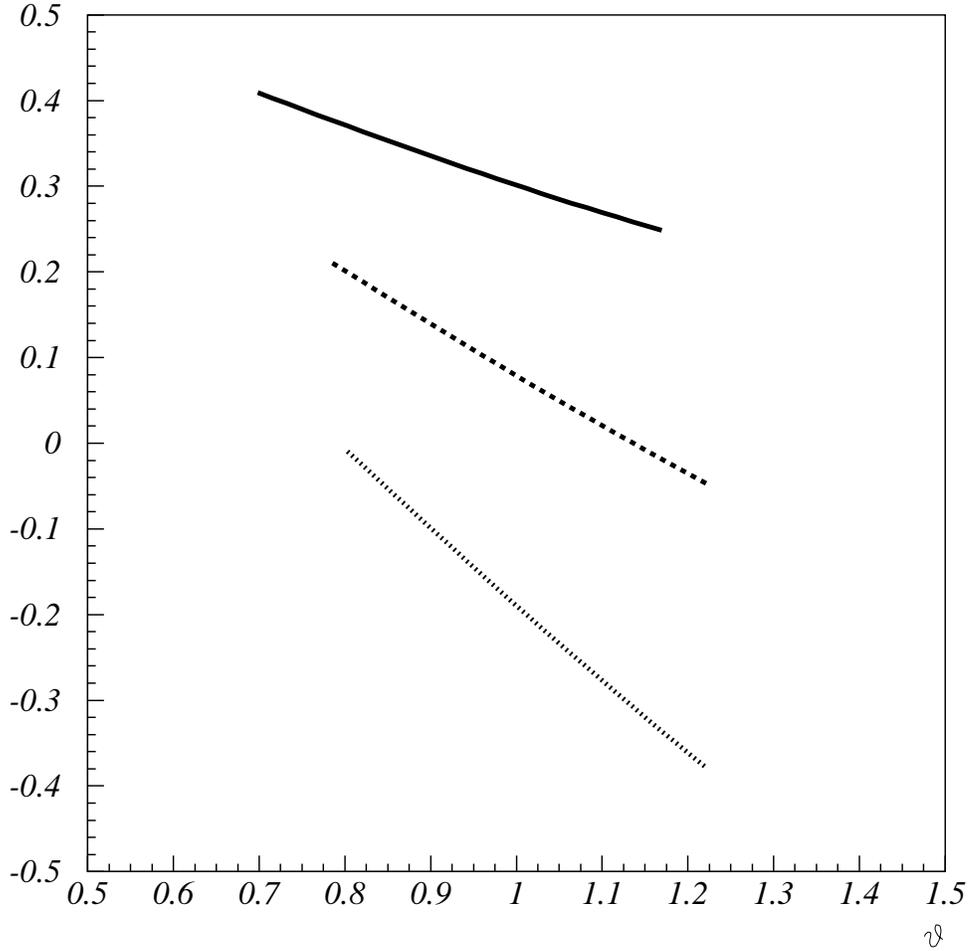}}
\caption{ $S_{\phi K}$
as a function of $\theta$ with
$\mu=2 m_b$ (solid), $m_b$ (dashed) and $m_b/2$ (dotted),
where $m_{H^{\pm}}=$ 200 GeV, $|\lambda_{tt}|=0.03$,
$|\lambda_{bb}|=100$, $\lambda_{ss}=\lambda_{cc}=100 e^{-i \pi/2}$.
Note that the masses of NHB [in Figs.
\ref{fig5}-\ref{fig11}] are different with those in
Figs. \ref{fig1}-\ref{fig4}.
}
\label{fig5}
\end{figure}

\begin{figure}
\epsfxsize=12 cm
\centerline{\epsffile{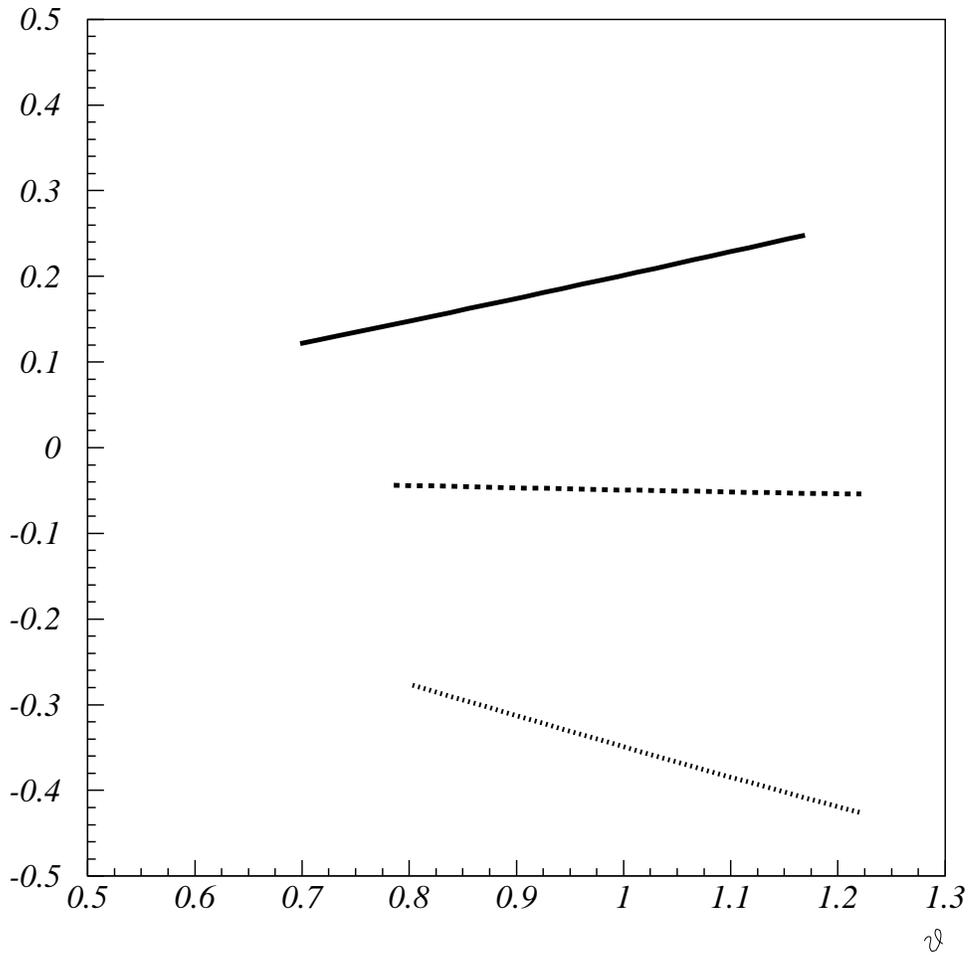}}
\caption{
$\Delta S$
as a function of $\theta$ with
$\mu=2 m_b$ (solid), $m_b$ (dashed) and $m_b/2$ (dotted).
 Other parameters are the same with Fig. \ref{fig5}.
}
\label{fig6}
\end{figure}

\begin{figure}
\epsfxsize=12 cm
\centerline{\epsffile{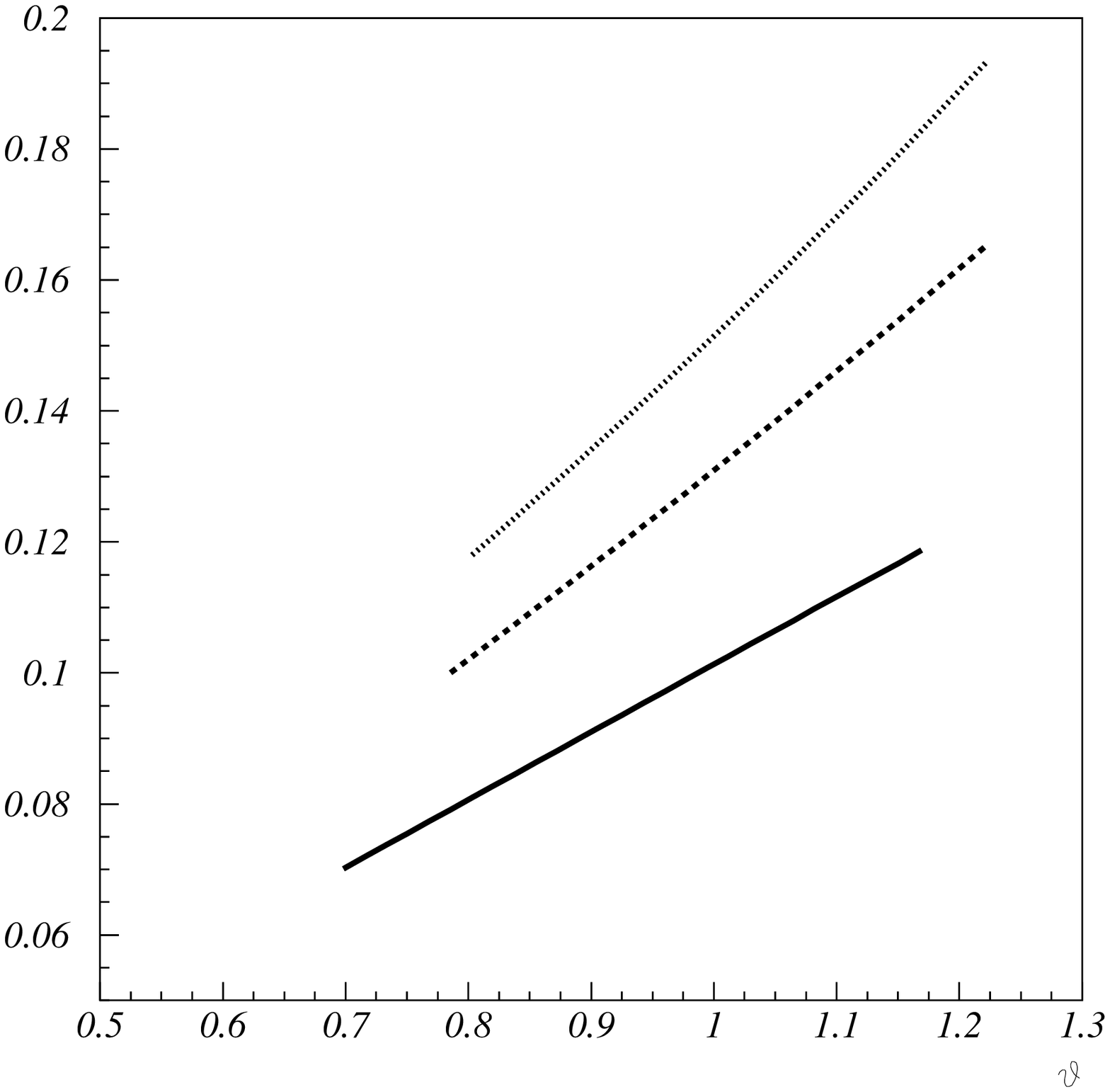}}
\caption{
$C_{\phi K}$ as a function of $\theta$. Other parameters and
conventions are the same with Fig. \ref{fig5}.}
\label{fig7}
\end{figure}

\begin{figure}
\epsfxsize=12 cm
\centerline{\epsffile{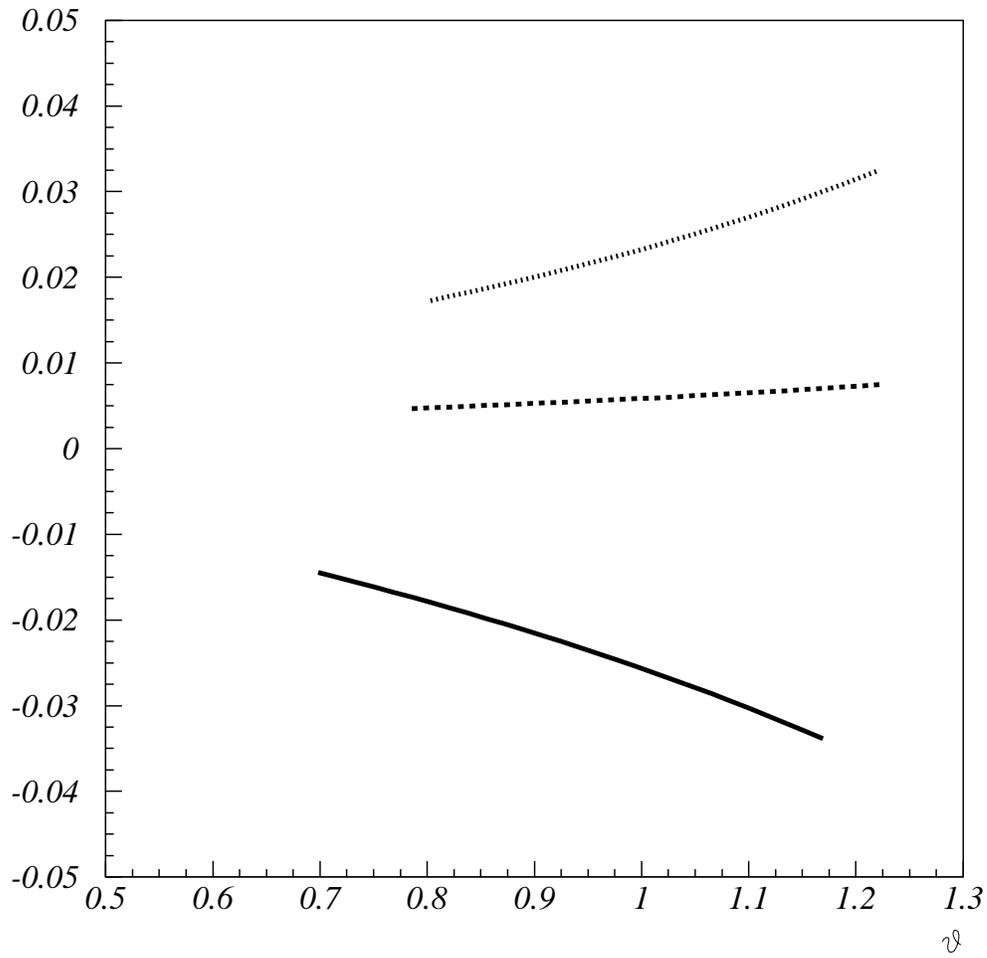}}
\caption{
$\Delta C$
as a function of $\theta$. Other parameters and
conventions are the same with Fig. \ref{fig7}.
}
\label{fig8}
\end{figure}

\begin{figure}
\epsfxsize=12 cm
\centerline{\epsffile{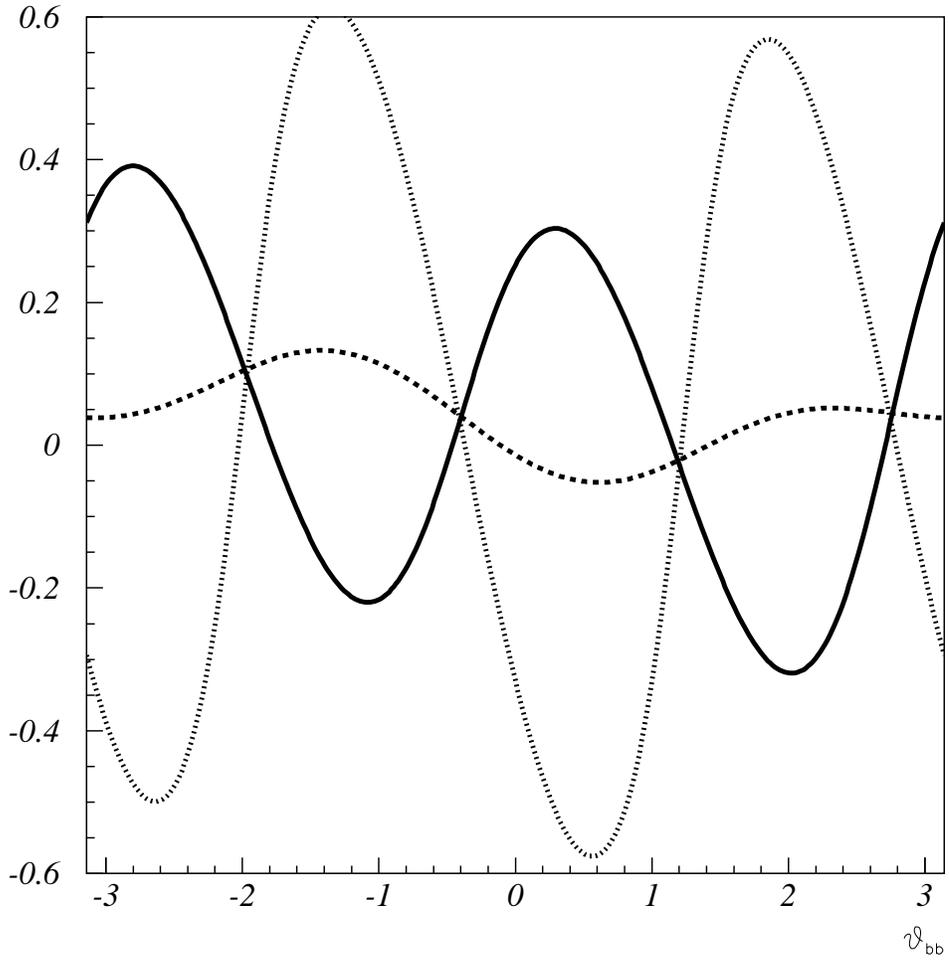}}
\caption{ $S_{\phi K}$
as a function of $\theta_{bb}$ with
$\mu=2 m_b$ (solid), $m_b$ (dashed) and $m_b/2$ (dotted),
where $m_{H^{\pm}}=$ 200 GeV, $|\lambda_{tt}|=0.03$, $|\lambda_{bb}|=100 $,
$\theta=1.15$ and
$\lambda_{cc}=\lambda_{ss}=100 e^{i \pi/4}$.
}
\label{fig9}
\end{figure}

\begin{figure}
\epsfxsize=12 cm
\centerline{\epsffile{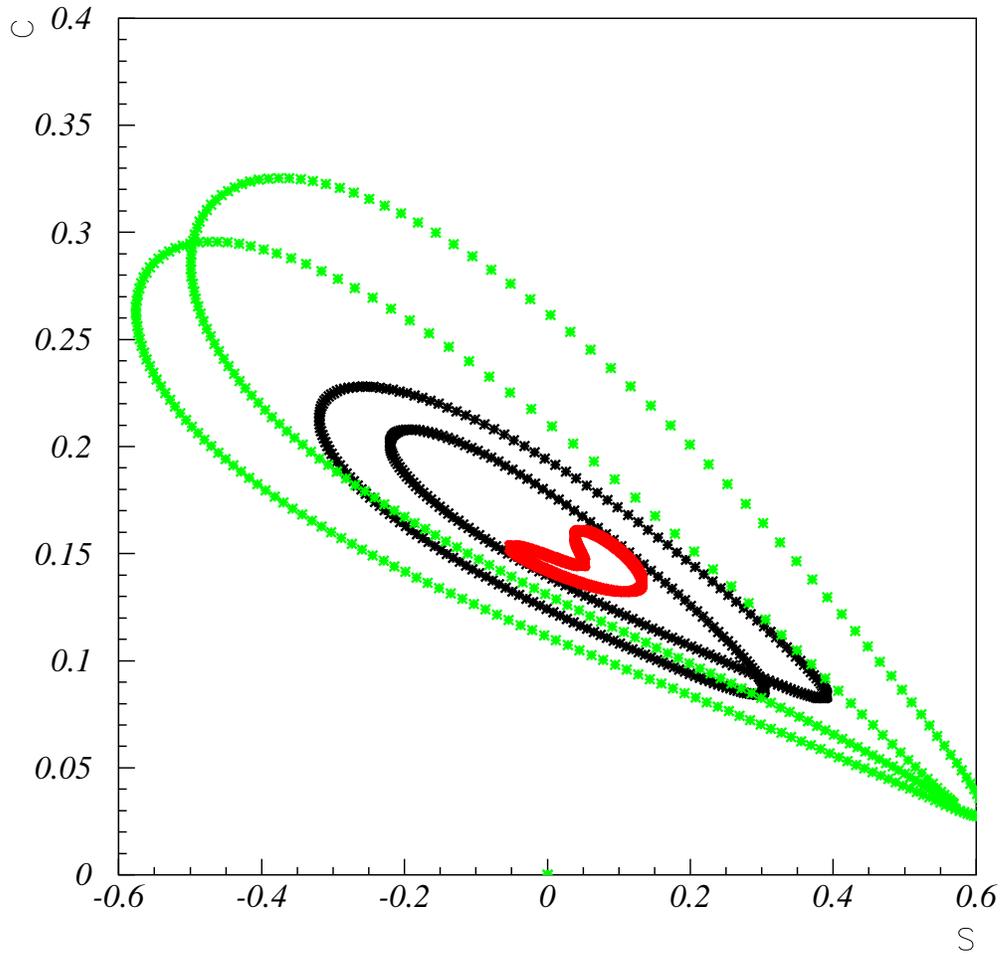}}
\caption{
Correlation between $C_{\phi K}$ and $S_{\phi K}$,
other parameters
are the same with Fig. \ref{fig9}. The outmost two curves correspond
to $\mu= m_b/2$, the curve in kernel is for $\mu= m_b$ and the other
two curves are for $\mu=  2 m_b$. }
\label{fig10}
\end{figure}

\begin{figure}
\epsfxsize=12 cm
\centerline{\epsffile{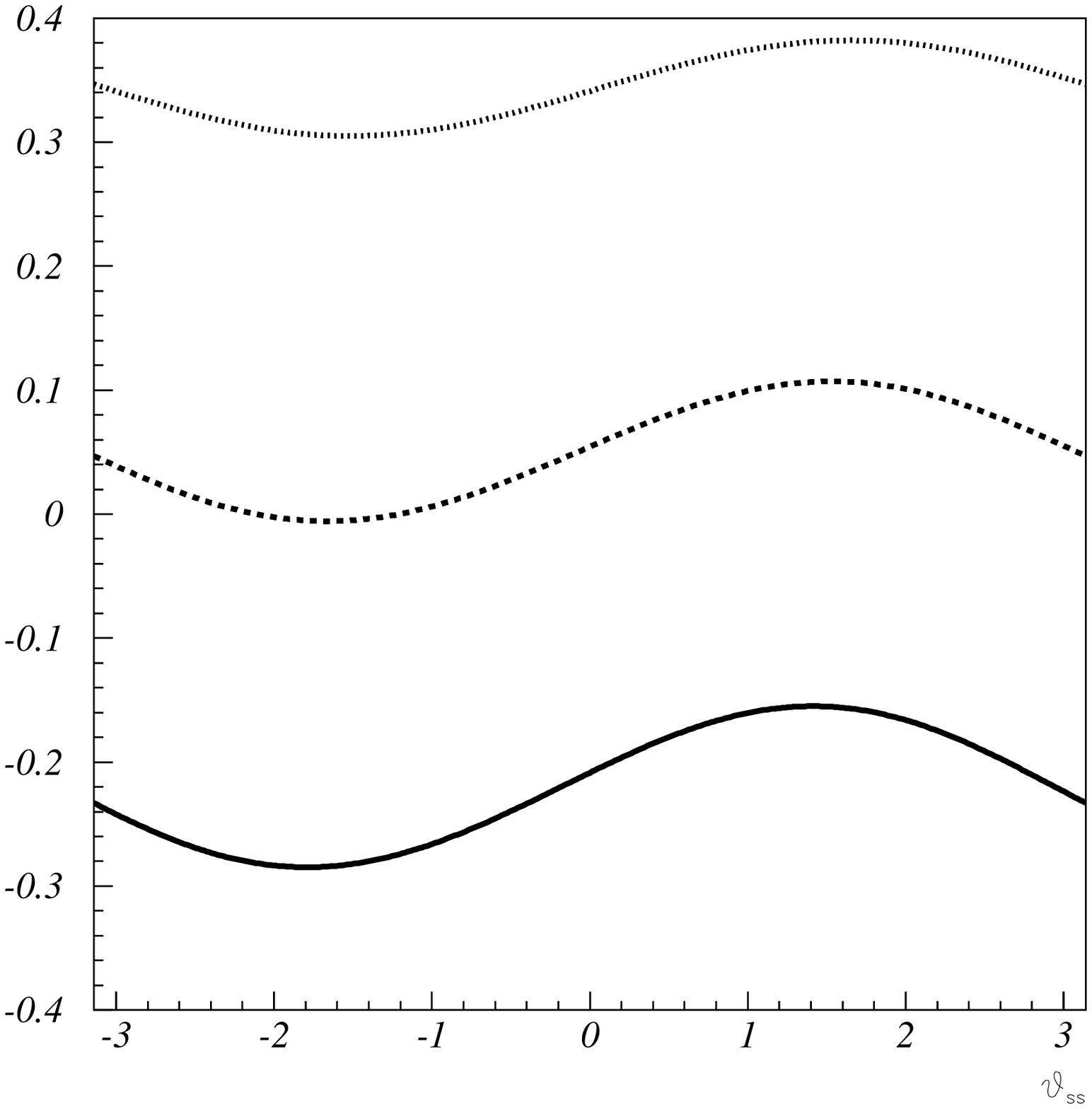}}
\caption{ $S_{\phi K}$
as a function of $\theta_{ss}$ with
$\mu=2 m_b$ (solid), $m_b$ (dashed) and $m_b/2$ (dotted),
where $m_{H^{\pm}}=$ 200 GeV, $|\lambda_{tt}|=0.03$, $|\lambda_{ss}|=100 $,
$\theta=1.15$,$\lambda_{bb}=100 e^{-i \pi/4}$ and
$\lambda_{cc}=100 e^{i \pi/4}$.
}
\label{fig11}
\end{figure}




\begin{thebibliography}{99}
\bibitem{newbelle} Belle
Collaboration, hep-ex/0308035.

\bibitem{newbabar} See talk given by T. Browder at LP2003, \\
{\em
http://conferences.fnal.gov/lp2003/program/S5/browder\_s05\_ungarbled.pdf.}


\bibitem{babarphi}B. Aubert {\it et al.}(BABAR Collaboration), hep-ex/0207070.

\bibitem{bellephi} T.~Augshev, talk given at ICHEP 2002 (Belle
Collaboration), BELLE-CONF-0232; K. Abe {\it et al.},
BELLE-CONF-0201 hep-ex/0207098; hep-ex/0212062.
\bibitem{mon}G. Hamel De Monchenault, hep-ex/0305055.
\bibitem{Anikeev:2001rk}
K.~Anikeev {\it et al.},
arXiv:hep-ph/0201071.
\bibitem{Grossman:1996ke}
Y.~Grossman and M.~P.~Worah,
Phys.\ Lett.\ B {\bf 395}, 241 (1997) [arXiv:hep-ph/9612269].

\bibitem{Fleischer:1996bv}
R.~Fleischer,
Int.\ J.\ Mod.\ Phys.\ A {\bf 12}, 2459 (1997)
[arXiv:hep-ph/9612446].



\bibitem{London:1997zk}
D.~London and A.~Soni,
Phys.\ Lett.\ B {\bf 407}, 61 (1997) [arXiv:hep-ph/9704277].

\bibitem{Grossman:1997gr}
Y.~Grossman, G.~Isidori and M.~P.~Worah,
Phys.\ Rev.\ D {\bf 58}, 057504 (1998) [arXiv:hep-ph/9708305].
\bibitem{Eigen:2001mk}
G.Eigen {\it et al.},
in {\it Proc. of Snowmass 2001},
hep-ph/0112312.
\bibitem{sphik}M.B. Causse, hep-ph/0207070; G. Hiller,
hep-ph/0207356; A. Datta, hep-ph/0208016; M. Ciuchini, L.
Silvestrini, hep-ph/0208087; S. Khalil and E. Kou, hep-ph/0212023;
G.L. Kane et al., hep-ph/0212092; R. Harnik, D.T. Larson and H.
Murayama, hep-ph/0212180; S. Khalil and E. Kou, hep-ph/0303214;
 K. Agashe and C.D. Carone, hep-ph/0304229; D.
Chakraverty et al., hep-ph/0306076; hep-ph/0307024;
A. Kundu and T. Mitra, Phys. Rev. {\bf D67}, 116005, (2003);
C.~W.~Chiang and J.~L.~Rosner,
Phys.\ Rev.\ D {\bf 68}, 014007 (2003) [arXiv:hep-ph/0302094].
\bibitem{chw}J.-F. Cheng, C.-S. Huang and X.-H. Wu,
hep-ph/0306086.

\bibitem{chao}D. Bowser-Chao, K. Cheung, and W.-Y. Keung, Phys.
Rev. {\bf D59} (1999) 115006 [arXiv:hep-ph/9811235]; Z.~j.~Xiao,
C.~S.~Li and K.~T.~Chao,
Phys.\ Rev.\ D {\bf 62}, 094008 (2000), ibid {\bf 63}, 074005
(2001), ibid {\bf 65} 114021 (2002);
Phys.\ Lett.\ B {\bf 473}, 148 (2000)


\bibitem{bbns}M. Beneke, G. Buchalla, M. Neubert and C.T. Sachrajda,
Nucl. Phys. {\bf B606}(2001) 245.

\bibitem{modelIII}
T.P. Cheng and M. Sher, Phys. Rev. {\bf D35}, 3484 (1987);
M. Sher and Y. Yuan, {\bf D44}, 1461 (1991); \\
W.S. Hou, Phys. Lett. {\bf B296} 179 (1992);\\
A. Antaramian, L. Hall, and A. Rasin, Phys. Rev. Lett. {\bf 69}, 1871 (1992);\\
L. Hall and S. Weinberg, Phys. Rev. {\bf D48}, 979 (1993);\\
M.J. Savage, Phys. Lett. {\bf B266}, 135 (1991); L. Wolfenstein
and Y.L. Wu, Phys. Rev. Lett. {\bf 73} (1994) 2809; D. Atwood, L.
Reina, and A. Soni, Phys. Rev. {\bf D55}, 3156 (1997).

\bibitem{Cvetic:1998uw}
G.~Cvetic, C.~S.~Kim and S.~S.~Hwang,
Phys.\ Rev.\ D {\bf 58}, 116003 (1998) [arXiv:hep-ph/9806282].



\bibitem{bbns1}M. Beneke, G. Buchalla, M. Neubert and C.T. Sachrajda, Phys. Rev. Lett. {\bf 83}(1999) 1914;
 Nucl. Phys. {\bf B591}(2000) 313.
\bibitem{dhll}Y.-B. Dai, C.-S. Huang, J.-T. Li, and W.-J. Li,
Phys. Rev. {\bf D67}, 096007 (2003).
\bibitem{burasbeast}
G.~Buchalla, A.~J.~Buras and M.~E.~Lautenbacher,
Rev.\ Mod.\ Phys.\  {\bf 68}, 1125 (1996)
[arXiv:hep-ph/9512380].

\bibitem{He:2000rf}
X.~G.~He, J.~P.~Ma and C.~Y.~Wu,
Phys.\ Rev.\ D {\bf 63}, 094004 (2001)
[arXiv:hep-ph/0008159].

\bibitem{Keum:2000ph}
Y.~Y.~Keum, H.~n.~Li and A.~I.~Sanda,
Phys.\ Lett.\ B {\bf 504}, 6 (2001) [arXiv:hep-ph/0004004].




\bibitem{adm}J.A. Bagger, K.T. Matchev and R.J. Zhang, Phys. Lett.
{\bf B412}(1997) 77; M. Ciuchini et al., Nucl. Phys. {\bf
B523}(1998) 501; C.-S. Huang and Q.-S. Yan, hep-ph/9906493; A.J.
Buras, M. Misiak and J. Urban, Nucl.Phys. {\bf B586} (2000) 397.




\bibitem{Kagan:1998bh}
A.~L.~Kagan and M.~Neubert,
Phys.\ Rev.\ D {\bf 58}, 094012 (1998)
[arXiv:hep-ph/9803368].
\bibitem{bu}I.I. Bigi and N.G. Uraltsev, Nucl. Phys. {\bf B353},
321 (1991).
\bibitem{gm}A.K. Giri and R. Mohanta, hep-ph/0306041.
\bibitem{27} G. Hiller and F. Krueger, hep-ph/0310219.
\bibitem{28} F. Borzumati et al., Phys. Rev. D62, 075005 (2000).



\end{thebibliography}
\end{document}